\documentclass[preprint]{article}
\usepackage{amsmath}
\usepackage{amssymb}
\usepackage{mathrsfs}
\usepackage[dvipdfmx]{graphicx}
\newtheorem{defn}{Definition}%[section]

\newtheorem{thm}[defn]{Theorem}
\newtheorem{lem}[defn]{Lemma}
\def\gam{\gamma}
\def\dh{\stackrel{h}{\nabla}}
\def\dgam{\stackrel{\gamma}{\nabla}}
\def\rh{\stackrel{h}{R}}
\def\rgam{\stackrel{\gamma}{R}}
\def\hoop{\operatorname{diam}}

\title{\bf\Large Upper Bound for Diameter of  Cosmological Black Holes
and Nonexistence of Black Strings}
\author{ Daisuke Ida\\
 Department of Physics, Gakushuin University,
 Tokyo 171-8588, Japan
}

\begin{document}
\maketitle
\begin{abstract}
The diameter of the apparent horizon, 
defined by the distance between furthest points on the horizon,
in spacetimes with a positive cosmological constant $\varLambda$
has been investigated. 
It is established that the diameter of the apparent horizon
on the totally umbilic partial Cauchy surface cannot exceed 
$2\pi/\sqrt{3\varLambda}$.
Then, it is argued that 
arbitrarily long 
black strings cannot be
formed in our universe.
\end{abstract}
\section{Introduction}

The general properties of black holes in classical general relativity have been extensively 
studied. In particular, the equilibrium problem of black holes in
asymptotically flat spacetimes is highlited by the uniqueness theorem for
the Kerr-Newman solution~\cite{heu96}.
The notion of black holes is extended to the cosmological setting,
where the cosmological horizon appears due to presence of the positive cosmological constant.
Although the equilibrium problem of such cosmological black holes 
is not established, 
they share a lot of beautiful properties with asymptotically flat black holes.
In particular,  many local results for the apparent horizon can be applied to
the cosmological black holes. For example, the Hawking's theorem ~\cite{haw72} that
asserts that the apparent horizon must be topological two-sphere 
also holds for black holes in spacetimes with a positive cosmological constant.

Cosmological black holes are fascinating in their own right.
They admit an interesting exact solution representing 
dynamical collision of black holes in the Einstein-Maxwell equations with the positive
cosmological term~\cite{kt93}.
As recent cosmological observations strongly suggest that our present universe
has a positive cosmological constant, it is very natural to
seek for the general properties of cosmologial black holes.
Hence, the main concern in this article is the black holes in four-dimensional spacetimes
with the positive cosmologcial constant $\varLambda$.

A remarkable property of the cosmological black hole is that the area of the  black hole
cannot exceed the value $4\pi/\varLambda$~\cite{snkm93,hsn94,mkni98}.
Thus, the black holes in inflationary universe cannot grow unboundedly, and so much 
large black holes cannot merge into one, 
or otherwise the naked singularity would be
formed~\cite{nsh95}.
One might however expect that more precise geometrical information about the black hole horizon 
would be obtained from the knowledge of
the appropriate length size of the black holes.
For example, the area bound does not controle the nonexistence of the black string solution,
as we can consider very thin and long horizons with the area of horizon fixed.

It is a general belief that there are no black string solution in four-dimensional general relativity.
This is supported by the absense of  known exact solutions
or  numerical examples.
A conclusive result excluding black strings, however, seems to be hardly known.

On the other hand, Thorne's hoop conjecture~\cite{tho72} in four-dimensilnal general relativity 
can be seen as an implication for the nonexistence of
such string-shaped black holes.
It claims that the black hole horizon forms if and only if the 
mass $M$ gets compacted into the region whose circumference $C$ in every direction
satisfies $C\le 4\pi M$.
Then, the only-if part of the conjecture claiming that a
realized horizon is subject to the above inequality seems to exclude 
arbitrarily long 
horizons for the given gravitational mass.
No counter example to the hoop conjecture has been reported,
while it has been tested for various exact solutions to the Einstein equation, or numerically generated spacetimes~\cite{nst88}.
Note however that we must appropriately define what is meant by mass, circumference, and horizon in the statement, when it is applied to the specific problem, since these notions are not specified there.

Nevertheless, the knowledge of
a characteristic length scale of the horizon, combined with 
that of its topology and area, would provide certain useful information about its geometric shape.
Here, we focus on the specific length scale of the apparent horizons in cosmological spacetimes that is given by the intrinsic distance between a furthest pair of points on the horizon,
which is proposed as a definition of half the circumference
in the Flanagan's work~\cite{fla91} seeking for the rigorous formulation of the hoop conjecture, and it is also known as the diameter of compact manifolds in differential geometry.

In order to analyze the diameter of the apparent horizon, we apply the
techniques of variational method in differential geometry,
which is developed
in the context of general relativity
 e.g. in Refs.~\cite{sy83,gm15}.

In the following note, we point out that the diameter of the apparent horizon of the cosmological black hole on the totally  umbilic partial Cauchy surface has the upper bound given by $2\pi/\sqrt{3\varLambda}$.
This seems to be the first conclusive example  that excludes the
existence of arbitrary long black strings in a certain class of 
cosmological spacetimes.

\section{The upper bound for the diameter of the black hole horizon}
Firstly, let us explain the general setting of the problem.
Let $M$ be the differentiable manifold endowed with the Lorentzian metric $g_{ab}$ with the signature $(-,+,+,+)$.
Let $\varSigma$ be a partial Cauchy surface in $M$, and let $U^a $ be a 
future-pointing timelike 
unit vector field on a neighborhood $\mathscr{U}$ of $\varSigma$, which is normal to $\varSigma$.
The tensor field 
\begin{align*}
  h_{a b }=g_{a b }+U_a  U_b 
\end{align*}
on $\mathscr{U}$ gives the Riemannian metric on $\varSigma$, when restricted to $\varSigma$.
Since $U^a $ is orthogonal to $\varSigma$, it satisfies
\begin{align*}
  U_{[a } \nabla_{b } U_{c]}=0
\end{align*}
on $\varSigma$. Then, the covariant derivative of $U_a $ is decomposed as
\begin{align*}
  \nabla_a  U_b =K_{a b }-U_a  A_b ,
\end{align*}
on $\varSigma$, where 
\begin{align*}
K_{a b }:=h_a ^c \nabla_c U_b   
\end{align*}
gives the second fundamental form of $\varSigma$, and
\begin{align*}
A_a :=U^b  \nabla_b  U_a 
\end{align*}
is the acceleration vector of $U^a $.
The restrictions of $K_{a b }$ and $A_a $ to $\varSigma$ are  tensor fields on $\varSigma$,
in the sense that these do not have a nonzero component tangent to $U^a $.

Let a closed 2-surface $H$ be an apparent horizon on $\varSigma$.
We consider a deformation of $H$ by 
\begin{align*}
  S: [-1/2,1/2]\times H\to \varSigma;
 (\xi,x)\longmapsto S_\xi (x),
\end{align*}
such that $S_0=i:H\hookrightarrow \varSigma$ is the inclusion map,
and that $S_\xi$ is a surface outside $H$ for $\xi>0$.

Let $N^a $ be the tangent vector field on $\operatorname{Im}(S)$,
which is the outward-pointing unit normal to $S_\xi$.
We define the tensor field on $\operatorname{Im}(S)$ as
\begin{align*}
  \gam_{a b }:=h_{a b }-N_a  N_b ,
\end{align*}
which gives the induced Riemannian metric on $S_\xi$.
The covariant derivative of $N_a$ is decomposed as
\begin{align*}
  \dh_a  N_b =\chi_{a b }+N_a  \alpha_b ,
\end{align*}
where
\begin{align*}
\chi_{a b }:=\gamma_a ^c \dh_c N_b 
\end{align*}
gives the second fundamental form of $H$ as a surface in $\varSigma$,
and
\begin{align*}
  \alpha_a :=N^b \dh_b  N_a 
\end{align*}
is defined.
The tensor fields $\chi_{a b }$ and $\alpha_a $ are regarded as those on $H$,
when they are restricted on $H$.
The normal vector field $N_a $ can be written as
\begin{align*}
  N_a =f \partial_a  \xi,
\end{align*}
where
the parameter $\xi$ of the deformation of $H$ is regarded as a function on
$\operatorname{Im}(S)$.
Then, it holds
\begin{align*}
  \alpha_a =-f^{-1}\partial_a  f.
\end{align*}

%%%%%%%%%%%%%%%%%%%%%%%%%%%%%%%%%%%%%%%%%%%%%
\begin{figure}[t]
%\centerline
\begin{center}
\includegraphics[width=.8\linewidth]{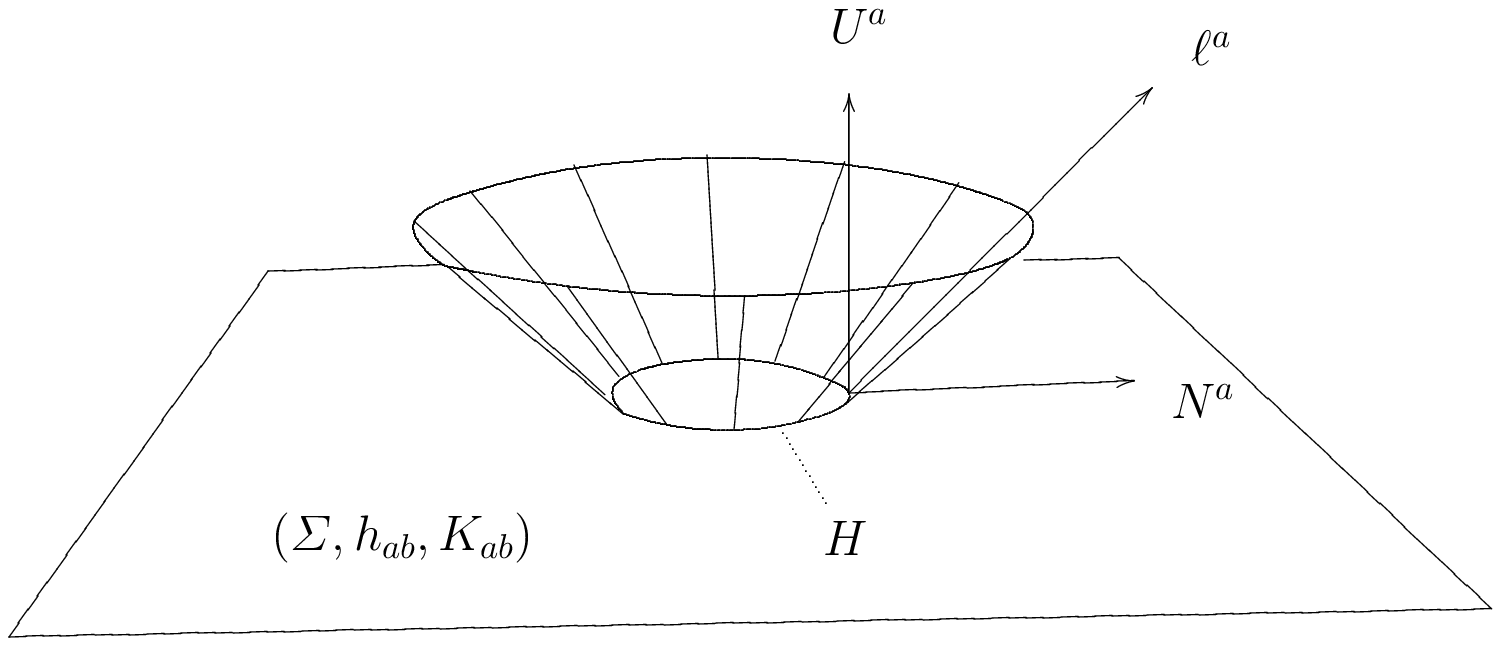}
\caption{A schematic picture representing the 
configuration of the apparent horizon $H$
on $\varSigma$.}\label{fig:fig02}
\end{center}
\end{figure}
%%%%%%%%%%%%%%%%%%%%%%%%%%%%%%%%%%%%%%%%%%%%%

The second fundamental form $K_{a b }$ of $\varSigma$ is decomposed as
\begin{align*}
  K_{a b }=\beta_{a b }+\zeta_a  N_b +N_a \zeta_b +\mu  N_a  N_b, 
\end{align*}
where
\begin{align*}
  \beta_{ab}&:=\gamma_a^c\gamma_b^dK_{cd},\\
\zeta_a&:=\gamma_a^b K_{bc}N^c,\\
\mu&:=K_{ab}N^aN^b
\end{align*}
give tensor fields on $H$.

The light rays in $M$ emanating from $S_\xi$ are tangent to the null vector field
\begin{align*}
  \ell^a:=U^a+N^a
\end{align*}
on $S_\xi$ (See Figure~\ref{fig:fig02}).
The expansion $\theta$ of the vector field $\ell^a$ is defined as
\begin{align*}
  \theta:=\gamma^{ab}\nabla_a \ell_b
=\beta+\chi,
\end{align*}
where we abbreviate
$\beta:=\beta^a_a$, $\chi:=\chi^a_a$.
The apparent horizon $H$ is a marginally trapped surface, i.e.,  it holds
\begin{align*}
  \theta=\beta+\chi=0,~~~\mbox{on $H$.}
\end{align*}

We write the vector field, which gives the deformation $S_\xi$ of $H$, as
\begin{align*}
  X^a =f N^a.
\end{align*}
Then, we obtain the differential of $\theta$ as
\begin{align}
\label{eq:dtheta}
f^{-1}  X^a \partial_a\theta=N^a\partial_a (\beta+\chi).
\end{align}

Here, using the Codazzi equation for $\varSigma$
\begin{align*}
  h_a^bR_{bc}U^c=\dh_b K_a^b-\dh_a K,
\end{align*}
we obtain
\begin{align*}
  R_{ab}N^aU^b&= N^a\dh_bK_a^b-N^a\dh_aK\\
&= \dgam_a\zeta^a-2\zeta_a\alpha^a+\mu\chi-\beta_{ab}\chi^{ab}-N^a\partial_a\beta,
\end{align*}
or
\begin{align}
\label{eq:Nbeta}
  \partial_N\beta=
\dgam_a\zeta^a-2\zeta_a\alpha^a+\mu\chi-\beta_{ab}\chi^{ab}-R_{ab}N^aU^b,
\end{align}
where $K:=K^a_a$ is defined and $\dgam_a$ denotes the covariant derivative on $H$.

From 
\begin{align*}
  \rh_{abcd}N^d&= (\dh_a\dh_b-\dh_b\dh_a)N_c\\
&= \dh_a\chi_{bc}-\dh_b\chi_{ac}+
N_a\alpha_b\alpha_c\\
&-\alpha_aN_b\alpha_c+
N_b\dh_a\alpha_c-N_a\dh_b\alpha_c,
\end{align*}
it follows that
\begin{align*}
  \rh_{ab}N^aN^b=-\chi_{ab}\chi^{ab}-\alpha_a\alpha^a+\dgam_a\alpha^a-N^a\partial_a\chi,
\end{align*}
or
\begin{align}
\label{eq:Nchi}
  \partial_N\chi=-\chi_{ab}\chi^{ab}-f^{-1}\dgam_a
\dgam
\stackrel{}{{}^a}
 f-\rh_{ab}N^aN^b
\end{align}
holds.

Using Eqs.~(\ref{eq:Nbeta}) and (\ref{eq:Nchi}), the Eq.~(\ref{eq:dtheta}) becomes 
\begin{align}
\label{eq:dtheta2}
  f^{-1}\partial_X\theta&=
\dgam_a\zeta^a-2\zeta_a\alpha^a+\mu\chi-\beta_{ab}\chi^{ab}-R_{ab}N^aU^b\nonumber\\
&-\chi_{ab}\chi^{ab}-f^{-1}\dgam_a
\dgam
\stackrel{}{{}^a}
 f-\rh_{ab}N^aN^b.
\end{align}

The Gauss equation for $\varSigma$
\begin{align*}
\rh_{abcd}=K_{ad}K_{bc}-K_{ac}K_{bd}+h_a^ph_b^qh_c^rh_d^sR_{pqrs}  
\end{align*}
leads to
\begin{align}
\label{eq:Rh}
 \rh=K_{ab}K^{ab} -K^2+R+2R_{ab}U^aU^b.
\end{align}
Also, the gauss equation for $H$ as a surface in $\varSigma$ 
\begin{align*}
  \rgam_{abcd}=\chi_{ac}\chi_{bd}-\chi_{ad}\chi_{bc}+h_a^ph_b^qh_c^rh_d^s\rh_{pqrs}  
\end{align*}
gives
\begin{align}
\label{eq:Rgam}
  \rgam=\chi^2-\chi_{ab}\chi^{ab}+\rh-2\rh_{ab}N^aN^b.
\end{align}
Eqs.~(\ref{eq:Rh}) and (\ref{eq:Rgam}) are put together into the form
\begin{align*}
2\rh_{ab}N^aN^b &= \chi^2-\chi_{ab}\chi^{ab}-\rgam\\
&+K_{ab}K^{ab} -K^2
+R+2R_{ab}U^aU^b\\
&= \chi^2-\chi_{ab}\chi^{ab}+\beta_{ab}\beta^{ab}+2\zeta_a\zeta^a-\beta^2-2\beta\mu\\
&-\rgam+R+2R_{ab}U^aU^b.
\end{align*}
Substituting this into Eq.~(\ref{eq:dtheta2}), we obtain
\begin{align}
\label{eq:dtheta3}
  f^{-1}\partial_X\theta
&=\dgam_a(\zeta^a-f^{-1}\dgam\stackrel{}{{}^a} f)
-(\zeta_a-f^{-1}\dgam_a f)(\zeta^a-f^{-1}\dgam\stackrel{}{{}^a} f)\nonumber\\
&-\dfrac{1}{2}\theta_{ab}\theta^{ab}
+\dfrac{1}{2}\theta^2+(\mu-\chi)\theta
+\dfrac{1}{2}\rgam
-8\pi G T_{ab}U^a\ell^b-\varLambda,
\end{align}
where we define
\begin{align*}
  \theta_{ab}:=\beta_{ab}+\chi_{ab},
\end{align*}
and the Einstein equation
\begin{align*}
  R_{ab}-\dfrac{1}{2}Rg_{ab}+\varLambda g_{ab}=8\pi G T_{ab}
\end{align*}
is applied.
Here and in what follows, we set the speed of light to unity.

For every deformation of $H$, which is determined by the positive function $f$
on $H$,  $S_\xi$ should not be a trapped surface for $\xi>0$, since $H$ is the
outermost trapped surface.
This requirement leads to the nonnegativity of the principal eigenvalue of the elliptic operator associated with Eq.~(\ref{eq:dtheta3})\footnote{The elliptic operator introduced here may not be a symmetric operator (i.e. with a drift term), so that its eigenvalues may be complex numbers. It however has the real eigenvalue $\lambda_1$, called the {\it principal eigenvalue}, such that $\lambda_1<\operatorname{Re}(\lambda)$ holds for every eigenvalue $\lambda\in\boldsymbol{C}$, and that the corresponding eigenfunction is a possitive function (See e.g. Ref.~\cite{eva98}, Chap. 6.).}.
\begin{lem}
\label{lem:1}
  Under the dominant energy condition, the principal eigenvalue of the linear operator
acting on the function on $H$:
  \begin{align*}
    A =
-\dgam_a\dgam\stackrel{}{{}^a} 
+2\zeta^a\dgam_a 
+\dfrac{1}{2}\rgam 
-\varLambda 
+(\dgam_a\zeta^a )
-\zeta_a\zeta^a
  \end{align*}
is nonnegative.
\end{lem}

\noindent {\it Proof.}
By definition, the dominant energy condition requires that
$T_{ab}V^aW^a\ge 0$ holds for every pair of future pointing timelike vectors $(V^a, W^a)$.
It follows that $T_{ab}U^a \ell^b\ge 0$ holds by continuity.

Let the real number $\lambda_1$ be the principal eivenvalue of $A$.
Consider the deformation of $H$ in terms of the deformation vector $X^a=fN^a$,
where the positive function $f$ is taken to be the corresponding eigenfunction $f$.
Then, on the apparent horizon $H$, the Eq.~(\ref{eq:dtheta3}) gives 
  \begin{align*}
\partial_X\theta&=
Af
-\dfrac{1}{2}\theta_{ab}\theta^{ab} f
-8\pi G T_{ab}U^a\ell^bf
\le \lambda_1 f.
  \end{align*}
It follows that $\lambda_1$ must be nonnegative, since otherwise we have $\partial_X \theta<0$ at every point on $H$,
to contradict to the condition that $H$ is the outermost trapped surface on $\varSigma$.
{\flushright$\Box$\par\medskip}

In the following, we consider a specific class of partial Cauchy surfaces given by
\begin{align*}
K_{ab}=\dfrac{1}{3}Kh_{ab},
\end{align*}
which we call the totally umbilic initial data $(\varSigma,h_{ab},K_{ab})$.
This restricted class of initial data is still allowed in
a wide class of spacetimes, such as 
the Kastor-Traschen multi-black-hole spacetimes~\cite{kt93}. 

Here, we show that a characteristic length of the horizon
must be not greater than the cosmological length scale, when $\varLambda$ is positive.
\begin{defn}
For a closed 2-surface $S$ in $\varSigma$, the diameter
of $S$ is defined by
\begin{align*}
\hoop(S):=\operatorname{max}\left\{\operatorname{dist}_S(p,q)|p,q\in S\right\},
\end{align*}
where $\operatorname{dist}_S(p,q)$ denotes the distance between $p$ and $q$ determined by the intrinsic geometry of $S$.
\end{defn}

\begin{thm}\label{thm:1}
  Let $(\varSigma,h_{ab},K_{ab})$ be a totally umbilic initial data
for the spacetime with the positive cosmological constant,
and let $H$ be the apparent horizon in $\varSigma$.
Under the dominant energy condition, the diameter
of $H$ satisfies
\begin{align}
\hoop(H)\le \dfrac{2\pi}{\sqrt{3\varLambda}}.
\end{align}
\end{thm}

\noindent {\it Proof.}
Take 
furthest pair of points $p$, $q$ on $H$.
Let $\varGamma:[0,L]\to H$ be the curve in $H$ connecting $p$ and $q$,
that minimizes the integral
\begin{align*}
  I_f:=\int_\varGamma f ds,
\end{align*}
where $f>0$ is the
principal eigenfunction of the linear operator $A$,
which in the present case $(\zeta^a=0)$ takes the form
\begin{align*}
  A=-\dgam_a\dgam\stackrel{}{{}^a} 
+\dfrac{1}{2}\rgam-\varLambda.
\end{align*}

Let $\nu^a$ be the unit vector field on the neighborhood of
$\varGamma$ in $H$, which is normal to $\varGamma$,
and let $g$ be a smooth real function on $\varGamma$ vanishing
at the endpoints $p$ and $q$.
Now we consider the variation of $I_f$ in terms of the
deformation vector $g\nu^a$.

The first variation of $I_f$ becomes
\begin{align*}
  \delta I_f=\int_\varGamma 
\left( \nu^a\partial_a f+f\dgam_a \nu^a\right)gds,
\end{align*}
so that 
\begin{align*}
\dgam_a \nu^a=- f^{-1} \partial_\nu f
\end{align*}
should hold on $\varGamma$.

Since $\varGamma$ mimizes $I_f$, its second variation should be
nonnegative.
This can be computed as
\begin{align*}
  \delta^2 I_f
&=\int_\varGamma
g\left\{
-f\dfrac{d^2g}{ds^2}
-\dfrac{df}{ds}\dfrac{dg}{ds}
+\left[\dgam_a \dgam\stackrel{}{{}^a} f
-\dfrac{\rgam}{2}f
-\dfrac{d^2f}{ds^2}
-f(\dgam_a\nu^a)^2
\right]g
\right\}
 ds.
\end{align*}
By Lemma~\ref{lem:1}, the inequality
\begin{align*}
  \dgam_a \dgam\stackrel{}{{}^a} f
-\dfrac{\rgam}{2}f\le-\varLambda f
\end{align*}
holds on $H$. Then, we have
\begin{align*}
\delta^2I_f&\le\int_\varGamma
\left(
-fg\dfrac{d^2g}{ds^2}
-g\dfrac{df}{ds}\dfrac{dg}{ds}
-g^2\dfrac{d^2f}{ds^2}
-\varLambda f g^2
\right)
 ds\\
&=
\int_\varGamma f
\left[
-2g\dfrac{d^2g}{ds^2}-\left(\dfrac{dg}{ds}\right)^2-\varLambda g^2
\right]ds.
\end{align*}
Now we take $g=[\sin(\pi s/ L)]^{2/3}$, where $L$ denotes the length of $\varGamma$.
Then, the above inequality leads to
\begin{align*}
\left(\dfrac{4\pi^2}{3L^2}-\varLambda\right)  \int_{\varGamma} fg^2 ds\ge 0.
\end{align*}
Hence, we have
\begin{align*}
  L\le \dfrac{2\pi}{\sqrt{3\varLambda}}.
\end{align*}
Since $\hoop(H)\le L$ holds by definition,
the statement of the theorem immediately follows.
{\flushright$\Box$\par\medskip}

Although Theorem~\ref{thm:1} does not make sense for $\varLambda=0$,
it is easy to obtain the version of Theorem~\ref{thm:1} without the
cosmological term,
by slightly modifying the above proof.
\begin{thm}\label{thm:2}
  Let $(\varSigma,h_{ab},K_{ab})$ be a totally umbilic initial data
for the Einstein equation
\begin{align*}
  R_{ab}-\dfrac{1}{2}Rg_{ab}=8\pi G T_{ab}.
\end{align*}
Let $\rho=T_{ab}U^aU^b$ and $J_a=h_a^cT_{bc}U^b$ be the
energy density and the energy flux of the matter field, respectively.
If an apprent horizon $H$ on $\varSigma$ is located within the region in which
\begin{align*}
8\pi G(  \rho-\sqrt{J_aJ^a})>c
\end{align*}
holds for a positive constant $c$,
then, the diameter
of $H$ satisfies
\begin{align*}
\hoop(H)< \dfrac{2\pi}{\sqrt{3c}}.
\end{align*}
\end{thm}

\noindent{\it Sketch of a Proof.}
This can be proved along similar lines to the reasoning of Lemma~\ref{lem:1} and Theorem~\ref{thm:1}, noting that
the condition on the energy current 4-vector implies that
the inequality
\begin{align*}
8\pi G T_{ab}U^a\ell^b>c
\end{align*}
holds on $H$, so that the linear operator
\begin{align*}
-\dgam_a \dgam\stackrel{}{{}^a} 
+\dfrac{1}{2}\rgam-c
\end{align*}
acting on the function on $H$
has the positive 
principal eigenvalue.
{\flushright$\Box$\par\medskip}

\section{Final Remarks}

We consider the apparent horizon
in spacetimes with a cosmological constant.
Then, we show that the diameter of the horizon 
on the totally umbilic partial Cauchy surface
has the upper bound given by $2\pi/\sqrt{3\varLambda}$ in terms
of the standard variational technique.
Since this upper bound depends only on the cosmological
constant,
it suggests the absense of 
arbitrarily long 
black strings
in the universe with a cosmological constant.

Though Theorem~\ref{thm:1} 
puts restrict on the hoop length of the black hole horizons,
it is not relevant for the Thorne's hoop conjecture.
In fact, the hoop conjecture with just that could tell
nothing about the arbitrary long black strings,
since it contains the gravitational mass scale in the
inequality.

It would be better 
if the condition of the total umbilicity in Theorem~\ref{thm:1}
could be relaxed,
since it is far from trivial if generic cosmological spacetimes
admit such a time slicing.

It is also unclear whether the present diameter bound is the best one or not.
Regarding the exact solutions, 
the supremum for the diameter of the apparent horizons
among the Schwarzschild-de Sitter class
is given by $\pi/\sqrt{\varLambda}$, which is nearly $87\%$ of
 $2\pi/\sqrt{3\varLambda}$.
As one direction of the future work, it might be interesting
to test the sharpness of the present diameter bound 
in terms of the numerical search 
of the apparent horizons for various initial data sets.

\end{document}